# High-performance chemical-bath deposited CdS thin-film transistors with ZrO$_2$ gate dielectric

Hareesh Dondapati, Duc Ha, Erin Jenrette, Bo Xiao and A. K. Pradhan*

Center for Materials Research, Norfolk State University, 700 Park Avenue, Norfolk, VA 23504, USA

*email address: apradhan@nsu.edu

**Abstract:**

We demonstrate high performance chemical bath deposited CdS thin-film transistors (TFTs) using atomic layer deposited ZrO$_2$ based high-k gate dielectric material. Our unique way of isolation of the CdS-based TFTs devices yielded significantly low leakage current as well as remarkable lower operating voltages (<5V) which is four times smaller than the devices reported on CdS-based TFTs using SiO$_2$ gate dielectric. Upon thermal annealing the devices demonstrate even higher performance, including $\mu_{FE}$ exceeding $4 \pm 0.2$ cm$^2$ V$^{-1}$S$^{-1}$, threshold voltage V$_T$ of 3.8V and I$_{on\text{-}off}$ of 10$^4$ to 10$^5$, which hold much promise for applications in future electronic and optical devices.

**Introduction**

Thin-film transistors (TFTs) are logical building blocks for diverse device applications such as flat, flexible, and transparent liquid crystal displays (LCDs).[1,2] Recent research on the development of high performance TFTs primarily focuses on its device limitations such as low mobility, ON/OFF ratio and fabrication cost.[3] II-VI semiconductors such as ZnO, CdS, PbS, PbSe and CdSe are expected to serve as promising materials for active channel layer in the development of high-performance TFTs for next-generation displays, alternative to conventional a-Si:H due to their higher electron mobilities and low processing cost.[4,5] Besides their attractive electrical and optical properties[6], such as bandgap tunability, and multiple exciton generation, inexpensive chemical-based low temperature fabrication techniques can be advantageous for mass production of large scale electronic devices at low cost. CdS is n-type semiconductor with a bandgap of 2.4 eV ( in bulk ), and it has been thoroughly explored as a window material in high efficiency thin film solar cells based on CdTe and Cu(In,Ga)Se$_2$ (CIGS).[7] However, there have been few reports on CdS TFTs fabricated using SiO$_2$ gate dielectrics, which demonstrate potential applications in display device technologies. CdS can be conveniently deposited uniformly using relatively very simple chemical bath deposition (CBD) technique for large scale production.[8] Despite significant research efforts invested in the development of CdS based TFTs, only limited reports are available on carrier mobility and operational stability. The CdS TFTs fabricated to date using conventional SiO$_2$ exhibit poor performance.[9] For instance, their on-to-off current ratios[10,11] and field-effect mobilities[12] are on the order of $10^2$ and as low as 0.12–0.16 cm$^2$ V$^{-1}$ s$^{-1}$, respectively.

We report here on the superior performance of CdS based TFTs fabricated using ZrO$_2$ gate dielectric layer deposited by atomic layer deposition (ALD). We have exploited two quite different thin film deposition methods together - CBD and ALD, in order to create CdS/ZrO$_2$ inexpensive based TFT structure for display matrixes. Typical thicknesses of the channel layers and dielectric materials in TFTs reported are about 100 nm.[13,14,] ALD is a well-known technique for the deposition of high-k dielectrics, moreover it gives highly conformal, defect-free dielectric layers at relatively low temperatures.[15,16] Dielectric films grown by ALD have been used in the fabrication of high-performance polycrystalline ZnO-TFTs.[17,18] With a high dielectric constant of 25, $ZrO_2$ is a good candidate to produce gate dielectrics with low leakage and high capacitance density.

Although the chemical based fabrication of thin films has the advantages of low temperature process and low cost, there remains the problem of a poor 'ON/OFF' drain current ratio due to the poor pinch-off characteristics resulting from the current through the as-grown CdS active channel layer. As thermal annealing[19], in the presence of forming gas, has a greater impact on the transport properties of CdS active layer, we used annealed active channel layers which yield high-performance TFTs.

Common bottom gate CdS TFTs were fabricated on RCA cleaned highly doped p-Si (100) substrates (0.001-0.005 ohm-cm) using ALD grown $ZrO_2$ gate oxide. Fig. 1(a) shows the schematic cross-section of CdS TFT with bottom gate configuration. Fig. 1 (b) is the top view of CdS TFTs fabricated on $ZrO_2$/*p-Si*. The device fabrication starts with the growth of $ZrO_2$ by ALD. A Savannah 100 ALD system from Cambridge Nanotech was used to deposit $ZrO_2$ dielectric films. The $ZrO_2$ films were deposited at 200 °C using alternating exposures of Tetrakis Dimethyl Amido Zirconium (TDMAZr) and water ($H_2O$) vapor at a deposition rate of approximately 0.09 nm per cycle.[20] Nitrogen gas was used as carrier and inert gas for purging the chamber. Each deposition cycle lasted 21s, yielding a total deposition time of around 6 h for 1000 cycles. ALD cycle consisting of TDMAZr pulse (0.25 s in duration), purge of the reaction zone with $N_2$ (10 s), $H_2O$ pulse (0.015 s) and another purge (10 s) was repeated until a film of required thickness was obtained. Phillips X-pert PRO system was used to analyze structure and thickness of the ZrO2 films. The $ZrO_2$ thickness was estimated as 90 nm and further confirmed using the cross sectional FESEM image is shown in 5 (b). The dielectric constant of $ZrO_2$ was calculated from the parallel plate capacitor structures fabricated on glass substrate by sandwiching $ZrO_2$ films between Ti/Au/Ti and top Ti/Au metal contacts. The CdS active layer is then deposited on the $ZrO_2$ grown silicon wafer by CBD.

A brief recipe of the active CdS layer deposition is as follows. The substrates used to deposit films are both commercial Corning glass slides and $ZrO_2$/p-Si wafers. Prior to the deposition for optical measurements glass substrates were cleaned in an ultra-sonicator with trichloroethylene, acetone, methanol and deionized water sequentially. The cleaned substrates were dried with high purity $N_2$ gas. CdS film are deposited using a typical procedure reported else where[12]. All the chemicals used were of analytical grade and procured from sigma Aldrich, and were used without further purification. A total of 100 ml aqueous chemical bath was prepared by dissolving 0.05 M cadmium chloride, $CdCl_2$, 0.5 M sodium citrate, $(C_6H_5O_7Na_3)$, 0.5 M potassium hydroxide, (KOH), 5 ml of pH 10 buffer solution and 0.5 M thiourea, $(CS(NH_2)_2)$ in deionized water. The beaker with the reaction solution was immersed in a water bath at 70 °C and substrates were placed vertically in the bath. The bath solution was initially colorless and the color turned gradually from yellow to orange, as the growth proceeded. CdS film growth was carried out at this temperature for 1h without stirring. The thickness of the films was measured using Dektak profilometer, and found to be 55 nm after 1h deposition. To achieve the typical thickness[13], of around 110 nm, the bath solution was replaced with the same concentration of reaction solution and carried out the deposition at the same temperature for another 1h without stirring. After deposition, the CdS films were cleaned in an ultrasonic bath with methanol, rinsed with distilled water, and dried with $N_2$ gas. Similar to the CdS films deposited on glass substrates, CdS active layers were deposited on $ZrO_2$/p-Si wafers for TFT fabrication. To improve the electrical performance of TFTs, CdS channel layers were annealed in formic gas at 350 °C for 1h. Source and drain electrodes were patterned directly on the top of CdS films using photolithography and standard lift-off process then followed by Au/Ti (55 nm/5 nm) metal deposition using e-beam evaporation technique. The CdS TFTs were isolated by a wet-etching process using hydrochloric acid (HCl: $H_2O$=1:100) diluted in de-ionized water in order to avoid gate leakage currents as well as to define the channels.

The normalized absorption spectra were recorded as a function of wavelength for both as-deposited and annealed CdS layers, is shown in Fig. 2 (a). It is noted that the thermal treatment causes a small redshift in absorption. As shown in Fig. 2 (b), the optical band gap of the film was determined by extrapolating the steep absorption edge to intercept a constant value from a plot of $(\alpha h\nu)^2$ versus $h\nu$, where $\alpha$ and $h\nu$ are absorption coefficient and photon energy. The estimated band gap of as grown CdS films is 2.51 eV at room temperature, however after thermal annealing the sample at 350 °C for 1h in in formic gas caused a decrease in band gap to 2.43 eV. This obtained value is very close to the bulk (standard) value for hexagonal CdS at room temperature in the literature.

The microstructure and surface morphology of both $ZrO_2$ and CdS layers were investigated by atomic force microscopy (AFM). Two dimensional AFM image of $ZrO_2$ film grown at 200 °C on Si showed in Fig. 3(a). $ZrO_2$ has a smooth surface with a small grain size and the rms (root mean square) roughness value obtained was 2.78 nm. From the AFM images of as grown CdS shown in fig 3(b) and annealed CdS films shown 3 (c), it is observed that CBD gives the dense and uniform distribution of larger grains with well-defined boundaries throughout the surface. The surface roughness of the as grown films was estimated to be 4 nm, however upon thermal annealing the roughness increased to 4.7 nm which is due to the coalescence of grains.

In order to study the crystal structures the X-ray diffraction (XRD) studies have been carried out on both $ZrO_2$ and CdS films. The XRD patterns of as grown $ZrO_2$ on Si substrate is presented in Fig. 4 (a). According to the previous reports, crystal phases of $ZrO_2$ such as monoclinic, orthorhombic, cubic, and tetragonal have strong influence on growth temperatures. $ZrO_2$ prepared by the ALD are polycrystalline in nature. In this study, a peak observed at 30° corresponds to (101) plane of t-$ZrO_2$ and the other two peaks observed at 35.3° and 49.2° correspond to (200) and (022) planes of m-$ZrO_2$, respectively. XRD pattern of as grown CdS on $ZrO_2$/Si is shown in Fig. 4 (b). The diffraction peaks observed at 26.8° and 33.2° corresponding to the (0002), and (220) planes of the hexagonal crystalline phase of CdS, however upon thermal annealing as shown in Fig. 4 (c), the relative large intensity of the (0002) diffraction peak clearly indicates CdS film is textured along the c-direction. This could be due to re-crystallization process in the polycrystalline film during the annealing process.

Field emission scanning electron microscopy (FESEM) images of as grown CdS films shown in Fig. 5 (a). We notice that CdS film grown on $ZrO_2$/Si substrate consists of a well-compacted granular structure with uniform nanocrystalline grains. The film is crack-free without observing noticeable pinholes and colloidal precipitates. This observation might suggest that the growth proceeded in a molecule-by molecule fashion presumably due to the controlled release of $Cd^{2+}$ ion in the presence of a strong complex agent of sodium citrate. Therefore, the average cluster sizes of as-deposited film were estimated from different clusters within the film and found to be about 40-60 nm. FESEM image of annealed films (not shown here) has shown no significant changes in topology. The thickness of both $ZrO_2$ and CdS extracted from the cross sectional FESEM image is shown in 5 (b). It is noted that CdS films grown by CBD for 2 h-at 70 °C yielded about 120 nm and the $ZrO_2$ grown by ALD for 1000 cycles gave about 90 nm thick films, which is comparable to the reported one[4] with equivalent oxide thickness of 14 nm.

The transfer and output characteristics of all CdS TFTs were measured using Keithley 4200-SCS parameter analyzer connected to a probe station. Electrical characteristics of CdS TFTs fabricated with channel length (L) = 50 μm and width (W) = 520 μm are presented Fig. 6. Drain current ($I_{DS}$) – drain voltage ($V_{DS}$) characteristics of CdS TFTs fabricated from as grown CdS channels are showed in Fig 6(a). It is clearly seen that the poor electrical response due to the fact that the grain boundaries in as grown polycrystalline samples have strong impact on the electrical transport as they act like dispersion and trapping centers for the charge carriers, reducing the mobility. And hence we were unable to extract the mobility of the semiconductor. On the other hand Fig. 6(b) shows typical output characteristics of CdS TFTs fabricated from annealed CdS channels measured in the voltage range from 0 to 15 V, for several gate voltages ($V_{GS}$), from 0 to 15 V in steps of 3V. It is observed that $I_{ds}$ increases markedly with increase of $V_{DS}$ at a positive gate bias $V_{GS}$. This implies that the channel is n-type and electron carriers are generated by positive $V_{GS}$. In the linear operation region ($V_{DS} \ll V_{GS}$) the drain current increases as a function of the drain voltage according to the following equation (1):

$$I_{DS} = \mu C_{ZrO_2} \frac{W}{L}(V_{GS}-V_T)V_{DS} \qquad (1)$$

Where $C_{ZrO2}$ is the capacitance per unit area of gate dielectric layer and $V_T$ is the threshold of the device. In the saturation region ($V_{DS} > V_{GS} - V_T$) the drain current is given by

$$I_{DS} = \frac{W\mu_{sat}C_{ZrO_2}}{2L}(V_{GS}-V_T)^2 \qquad (2)$$

Where $\mu_{sat}$ is the carrier mobility. From the equation (2) the mobility and threshold voltage of the devices are extracted by linear fitting. The sqrt ($I_{DS}$) versus $V_{GS}$ curves shown in Fig 6 (c) were measured at a fixed $V_{DS}$ of 10 V. The threshold voltage and $I_{on}/I_{off}$ obtained from the plots in this figure are 3.8 V and $2\times10^4$, respectively, showing that the TFT operates in the enhancement mode. These characteristics are much improved over those reported for CdS TFTs fabricated using SiO$_2$. It is noted that TFTs made using HfO$_2$ as a gate dielectric[4] show comparable operating voltages as in our ZrO$_2$-based TFTs, however the leakage current in our case is significantly low, of the order of 3.2×10$^{-6}$ A/Cm$^2$ at 10V.

In summary, high performance CdS TFTs are fabricated using two low cost methods, such as CBD and ALD, and a detailed study of how thermal annealing influences its performance is presented. Well-compacted, highly uniform and crack free CdS thin films were deposited on various substrates using CBD technique at 70 °C. A detailed study of optical, structural, and morphological characterization of CdS thin films demonstrated superior film quality. We have achieved significantly low gate leakage current, in the order of 10$^{-9}$ A, with unique way of isolation of each device. We have demonstrated CdS-based TFTs fabricated using ZrO$_2$ high k dielectric material upon thermal annealing, which showed field effect mobility $\mu_{FE}$ exceeding 4 ± 0.2 cm$^2$ V$^{-1}$S$^{-1}$, and low operating voltages $V_T$ of 3.8V and $I_{on-off}$ of 10$^4$ to 10$^5$. Our findings suggest much promise for applications in future electronic and optical devices.

## Acknowledgements

This work is supported by the NSF-CREST (CNBMD) Grant No. HRD 1036494, and partially by DoD (CEAND) Grant No. W911NF-11-1-0209 (US Army Research Office). The authors are thankful to M. Bahoura for critically reading the manuscript. The authors would like to thank Brandon Walker for experimental help.

**Figure Captions**

Fig. 1(a) Schematic cross-section of CdS TFT in bottom gate configuration. (b) Top view of CdS TFTs fabricated on $ZrO_2/p$-Si.

Fig. 2(a) Absorption spectra (b) variation of band gap in CdS films deposited on glass substrates.

Fig. 3. AFM images of (a) $ZrO_2$ (b) as grown CdS and (c) annealed CdS on $ZrO_2$/Si substrate.

Fig. 4. XRD spectra of (a) $ZrO_2$/Si (b) as deposited CdS on $ZrO_2$/Si (c) annealed CdS on $ZrO_2$/Si.

Fig. 5. FESEM images (a) top view CdS and (b) cross sectional image of $ZrO_2$ and CdS on Si.

Fig. 6. Output characteristics of CdS TFTs fabricated from (a) as grown CdS and (b) annealed CdS films.

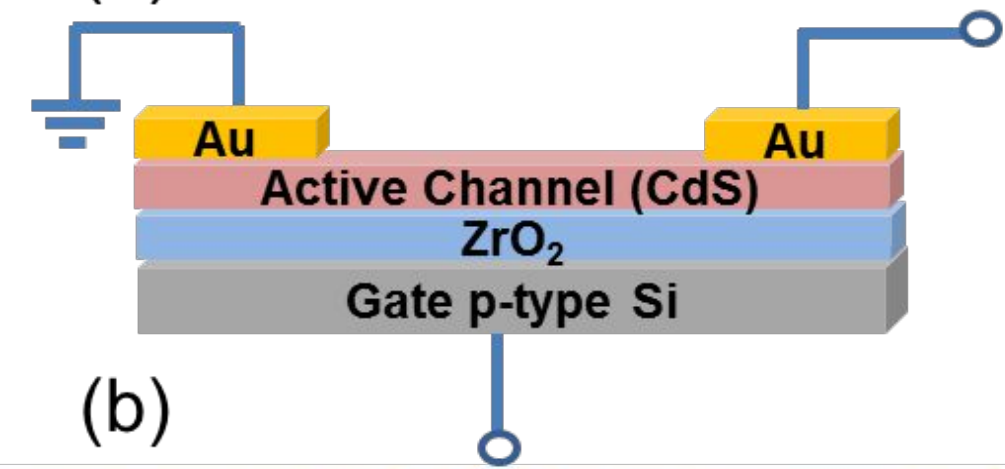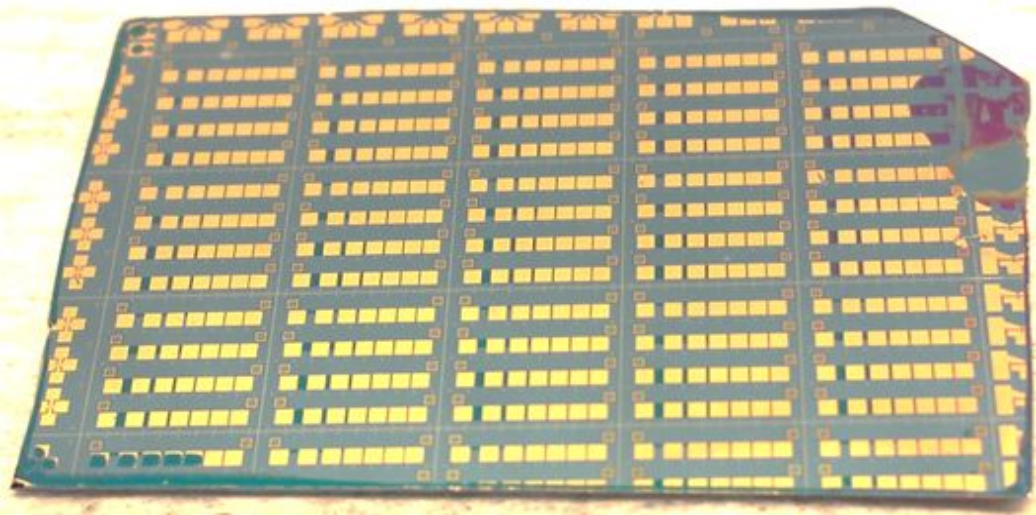

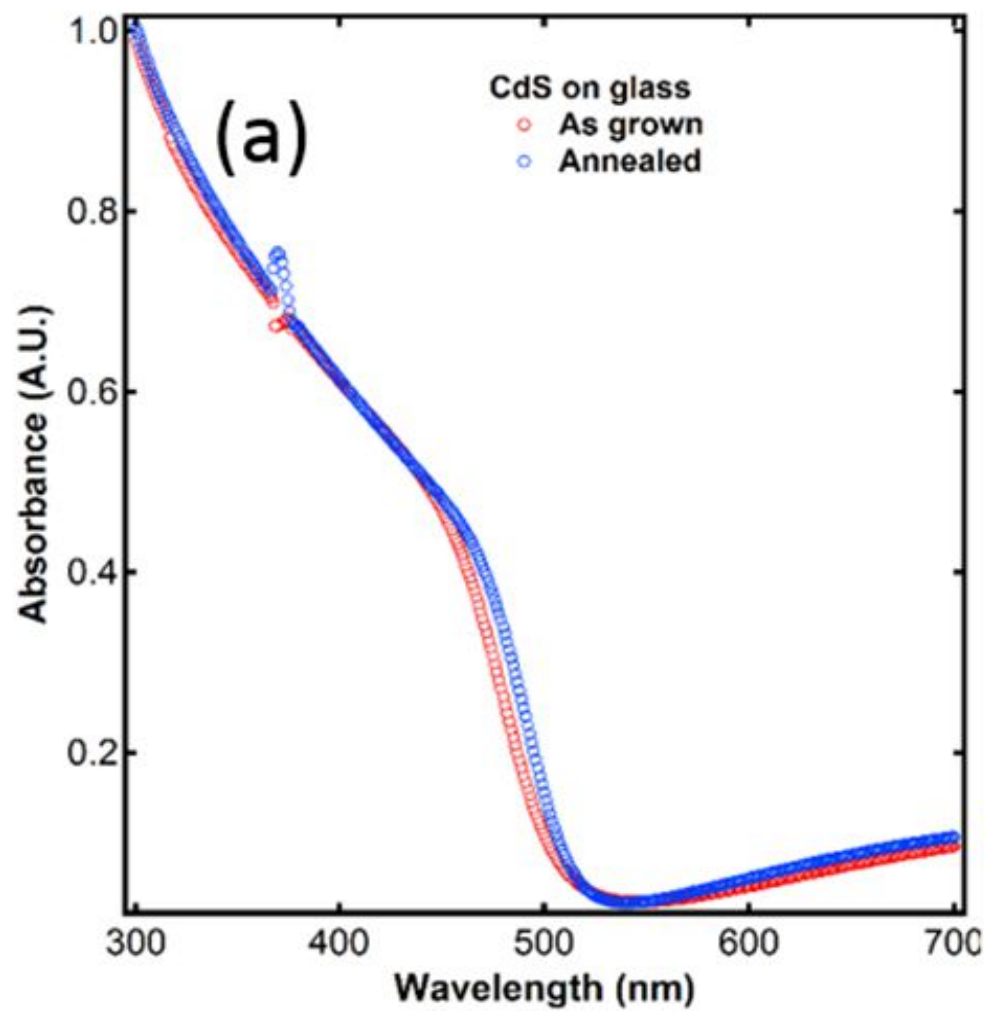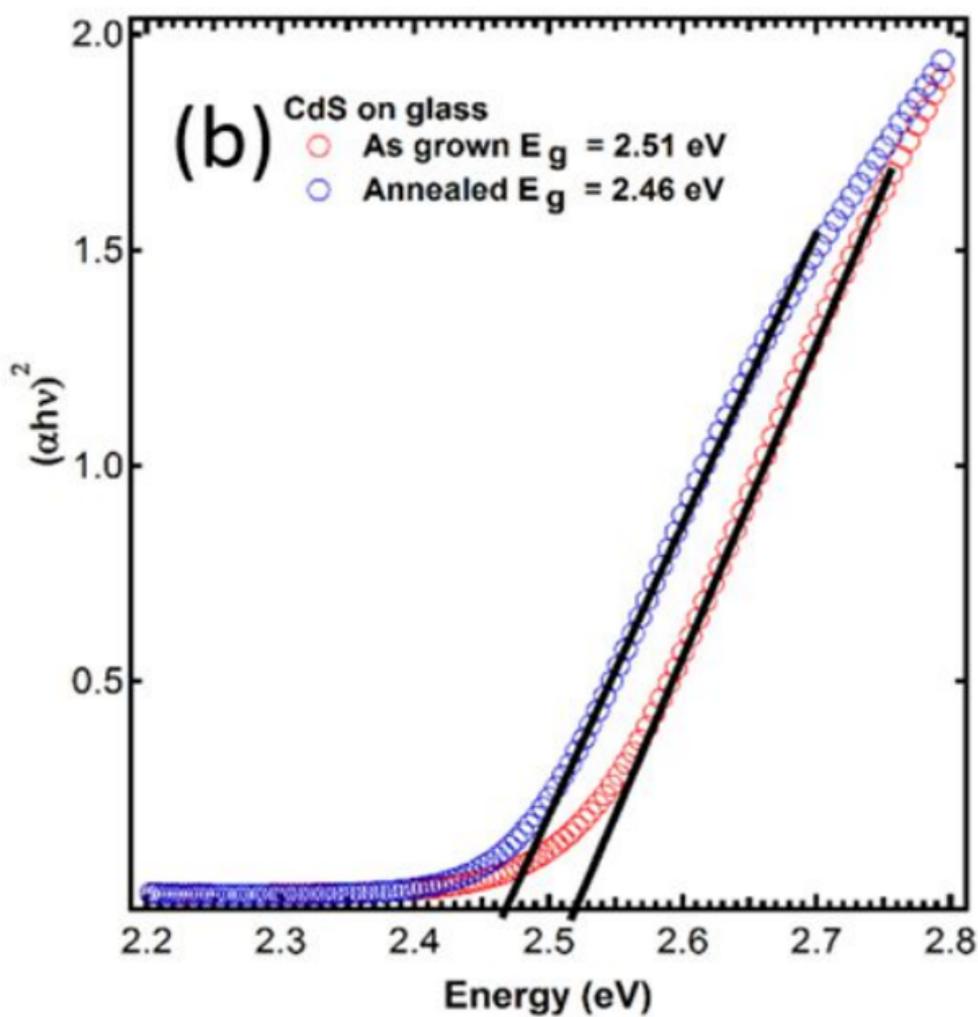

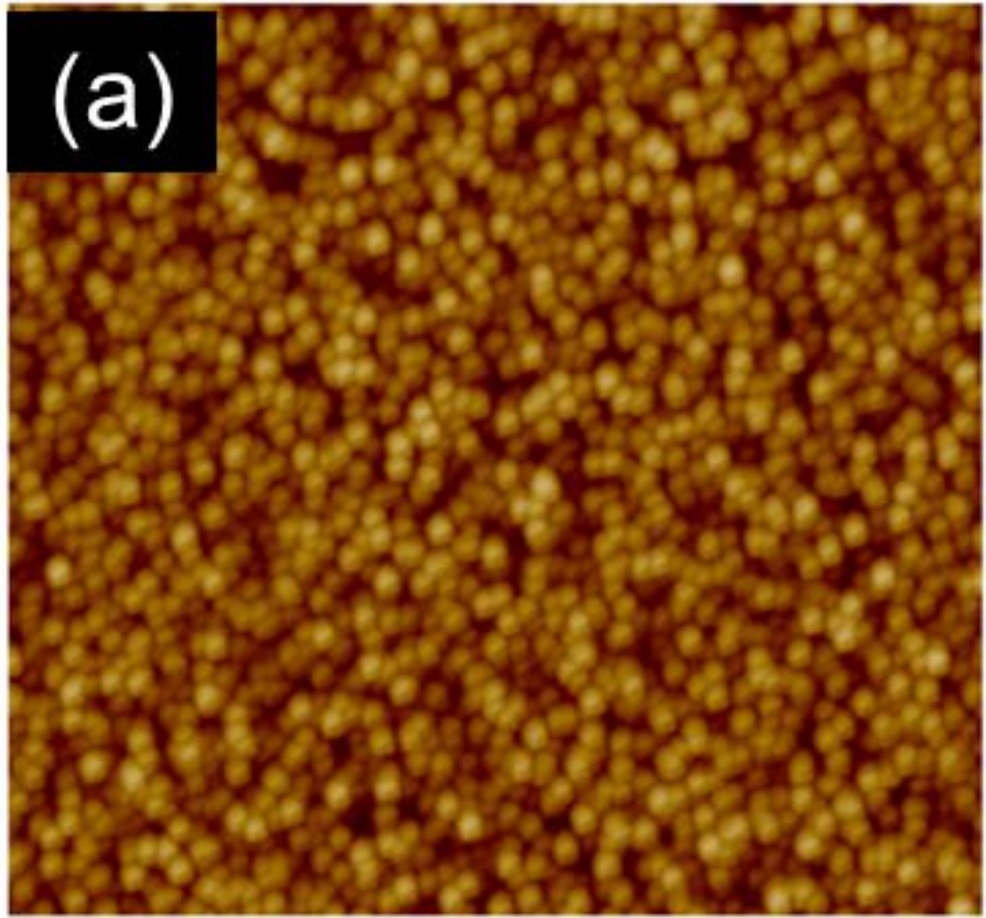
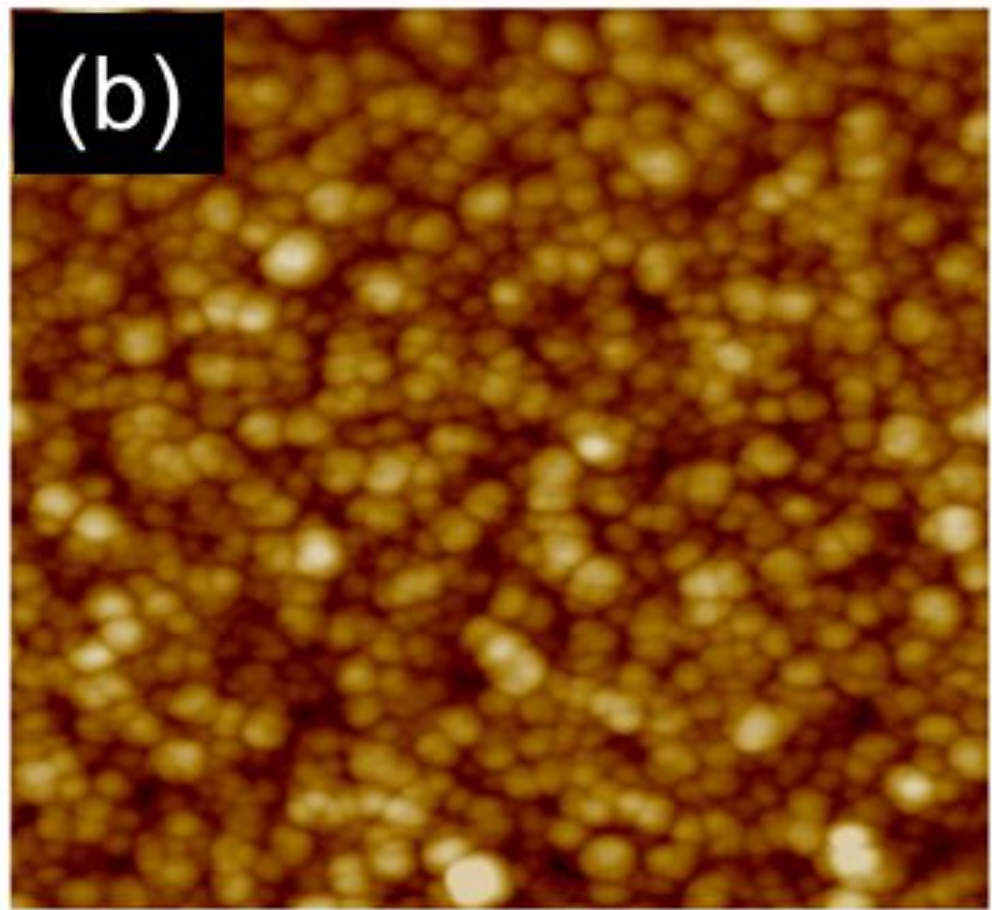
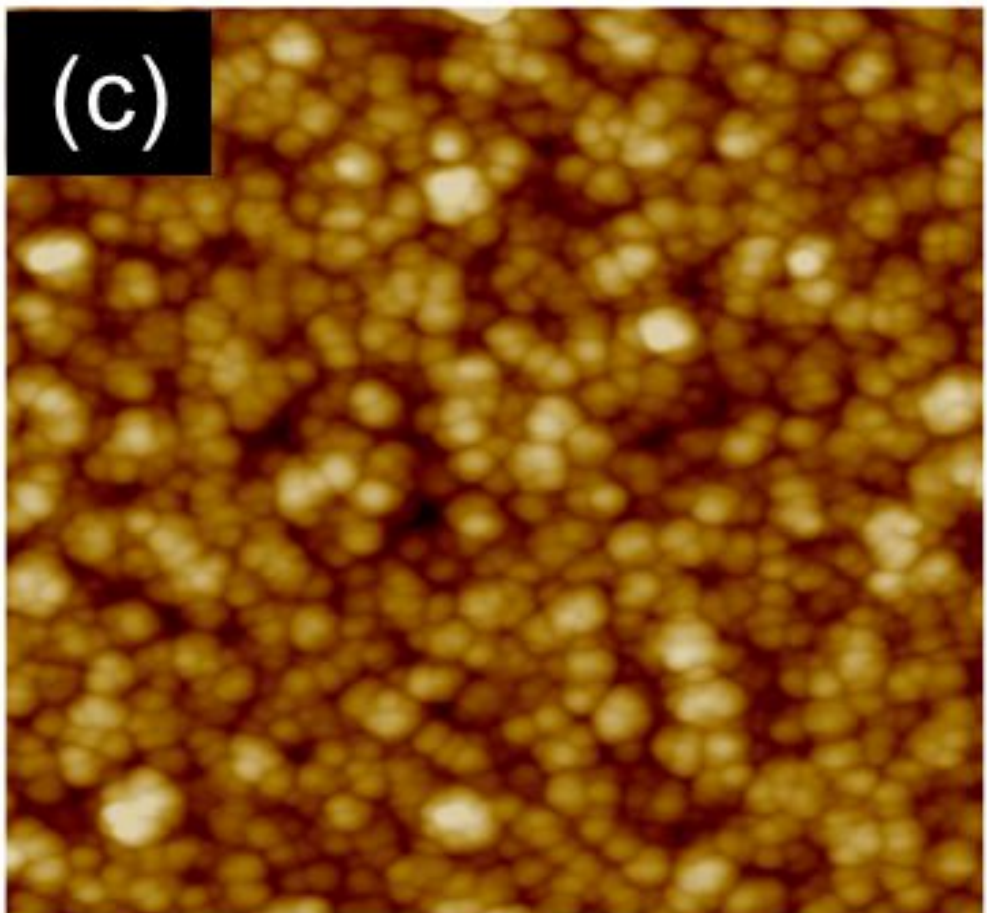

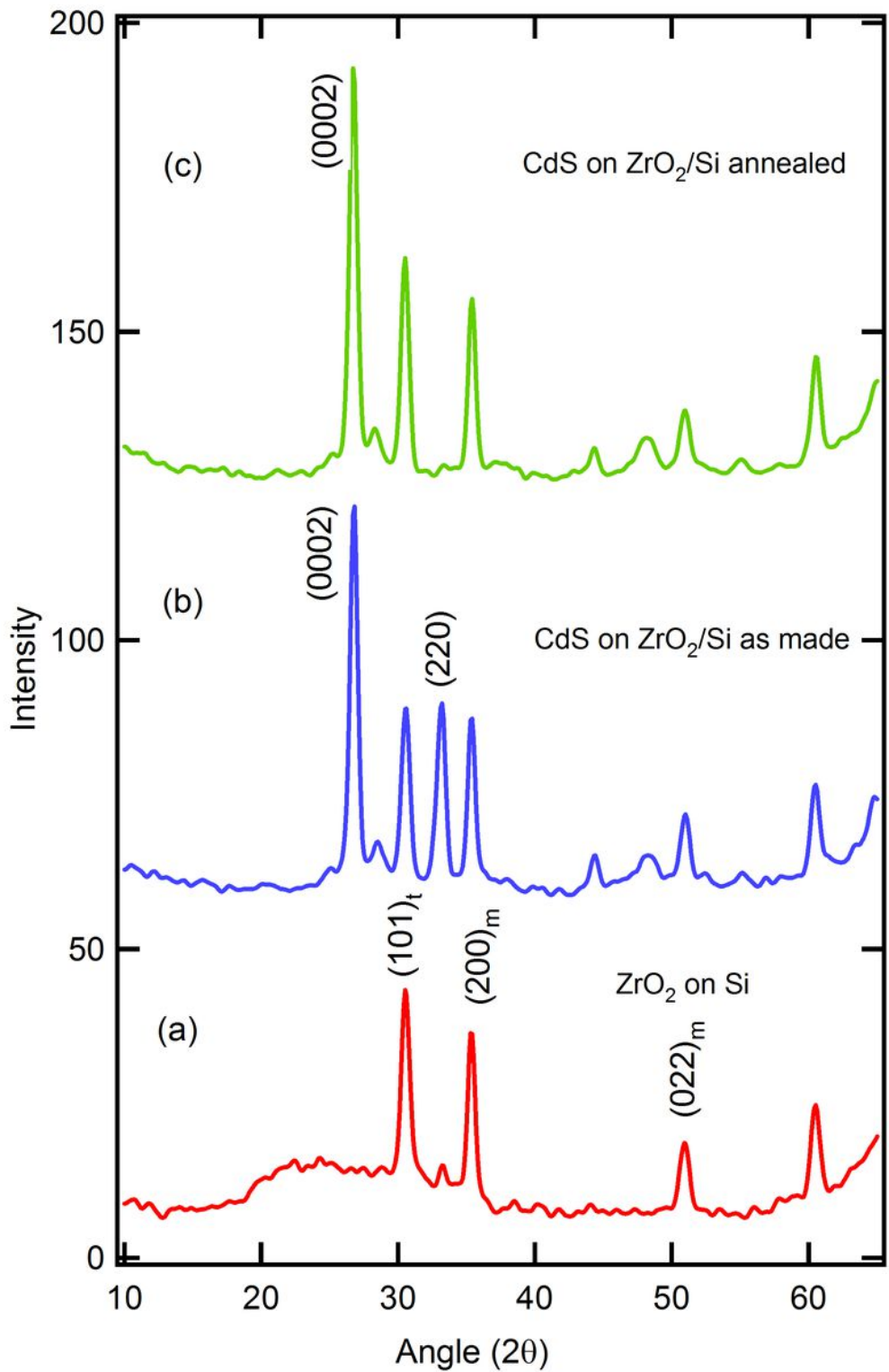

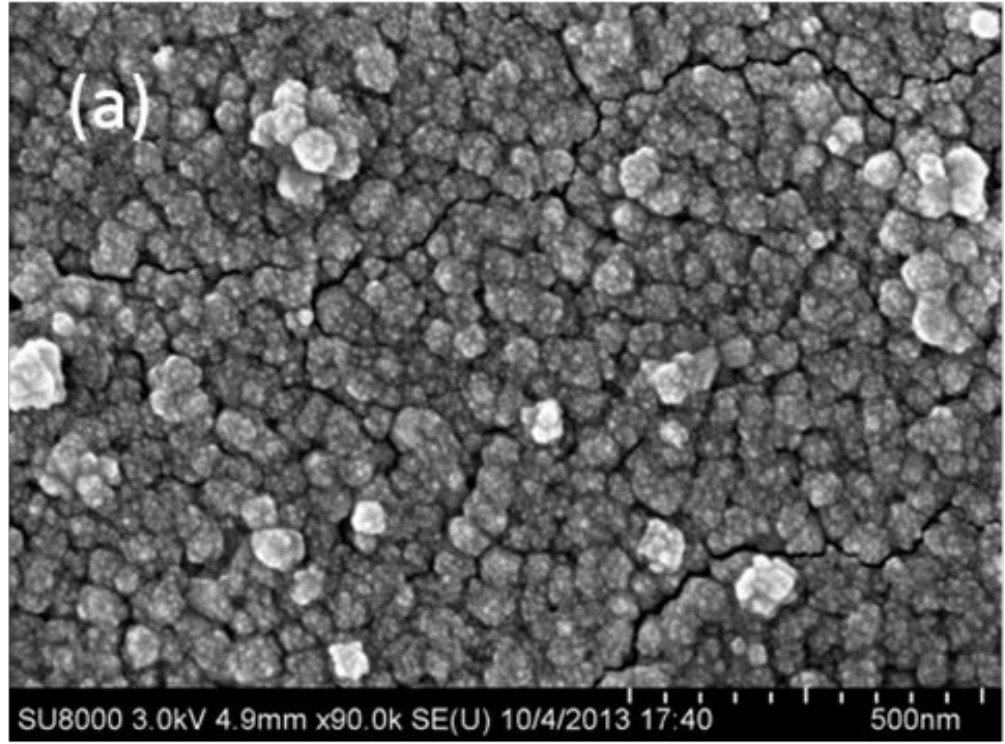

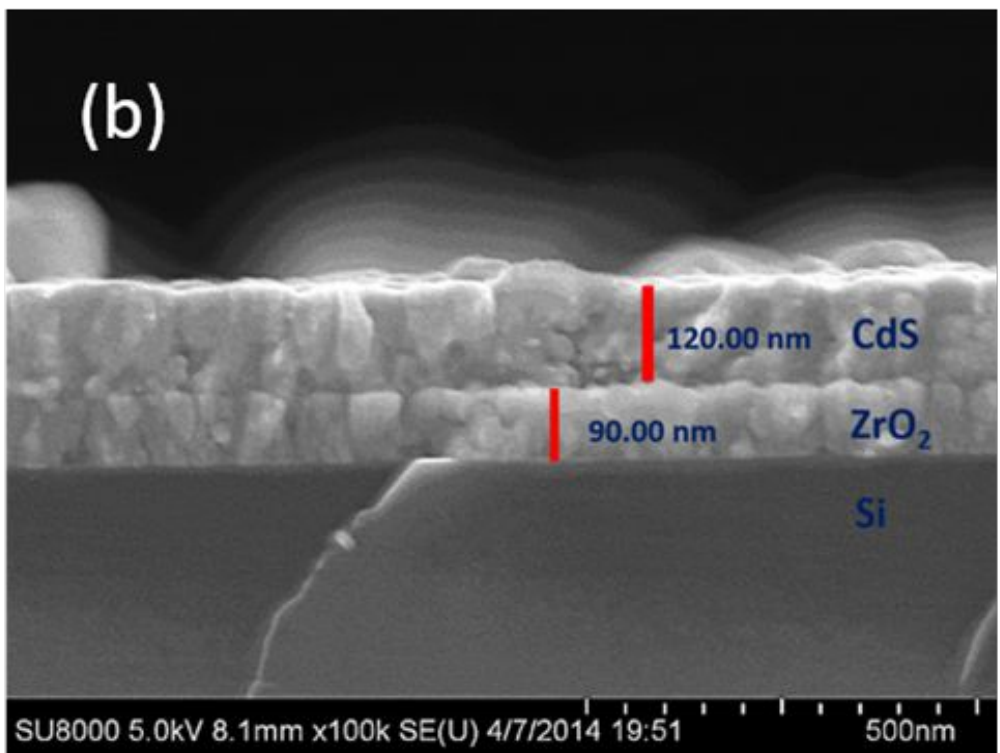

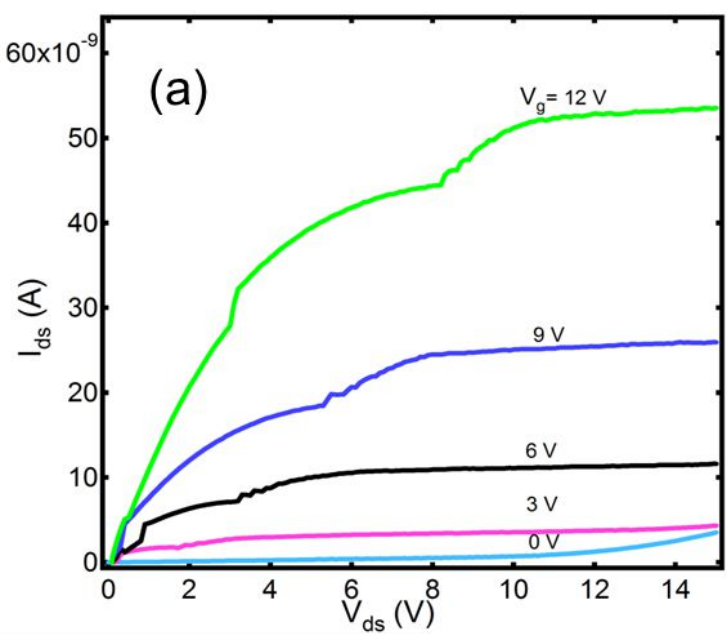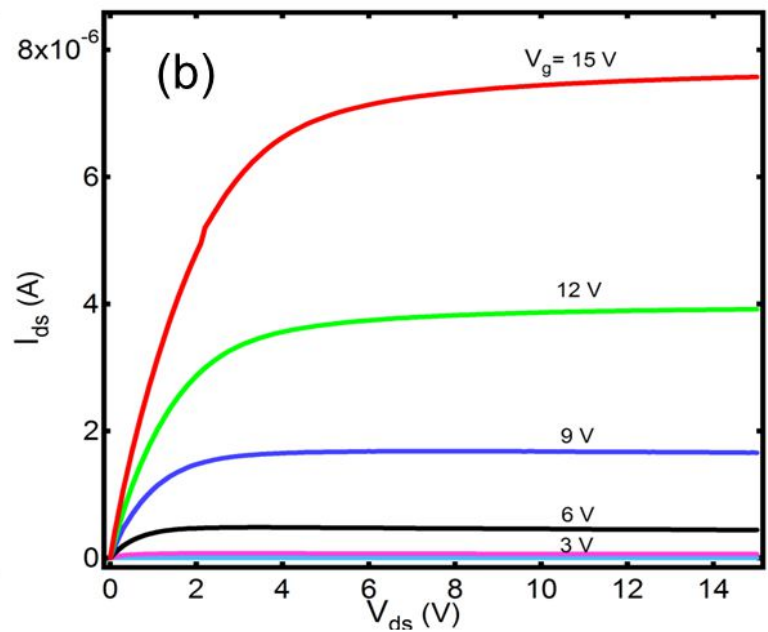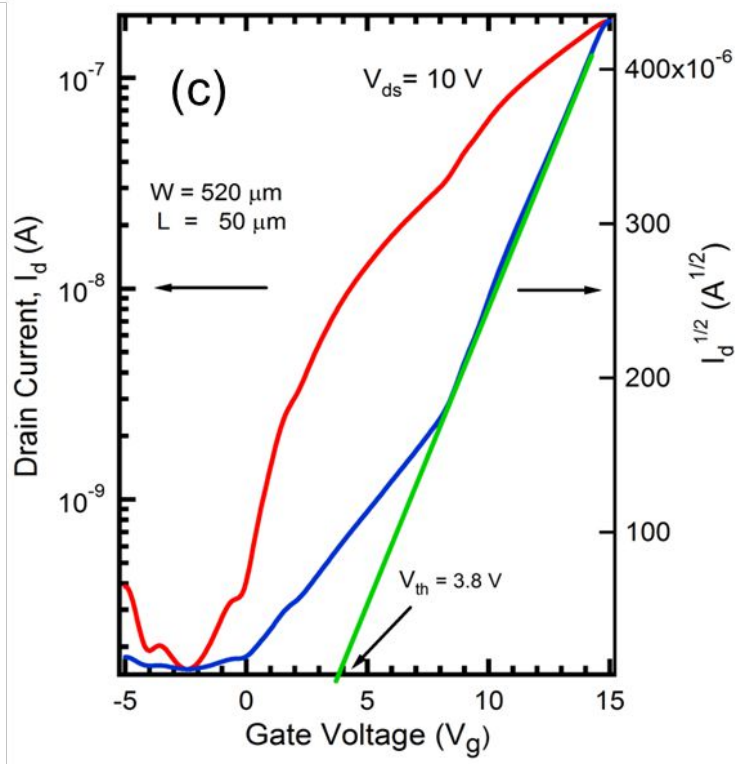